\documentclass[iop]{emulateapj}
\usepackage{color,hyperref}
\definecolor{linkcolor}{rgb}{0,0,0.5}
\definecolor{notecolor}{rgb}{0.8,0,0}
\hypersetup{colorlinks=true, linkcolor=linkcolor, citecolor=linkcolor, 
  filecolor=linkcolor, urlcolor=linkcolor}
\usepackage{amssymb,amsmath}
\usepackage{natbib}

\begin{document}

\title{Chemical constraints on the contribution of Population III
  stars to cosmic reionization} \author{Girish
  Kulkarni\altaffilmark{1}, Joseph F.~Hennawi\altaffilmark{1},
  Emmanuel Rollinde\altaffilmark{2}, Elisabeth
  Vangioni\altaffilmark{2} } \altaffiltext{1}{Max Planck Institute for
  Astronomy, K\"onigstuhl 17, 69117 Heidelberg, Germany;
  \email{girish@mpia-hd.mpg.de}} \altaffiltext{2}{Institut
  d'Astrophysique de Paris, UMR 7095, UPMC, Paris VI, 98 bis boulevard
  Arago, 75014 Paris, France}

\begin{abstract}
Recent studies have highlighted that galaxies at $z = 6$--$8$ fall
short of producing enough ionizing photons to reionize the IGM, and
suggest that Population~III stars could resolve this tension, because
their harder spectra can produce $\sim 10\times$ more ionizing photons
than Population~II.  But this argument depends critically on the
duration of the Population~III era, and because Population~III stars
form from pristine gas, in turn depends on the rate of galactic
enrichment.  We use a semi-analytic model of galaxy formation which
tracks galactic chemical evolution, to gauge the impact of
Population~III stars on reionization. Population~III SNe produce
distinct metal abundances, and we argue that the duration of the
Population~III era can be constrained by precise relative abundance
measurements in high-$z$ damped Ly$\alpha$ absorbers (DLAs), which
provide a chemical record of past star-formation. We find that a
single generation of Population~III stars can self-enrich galaxies
above the critical metallicity $Z_{\rm crit}=10^{-4}Z_\odot$ for the
Population~III-to-II transition, on a very short timescale $t_{\rm
  self-enrich}\sim 10^{6}$ yr, owing to the large metal yields and
short lifetimes of Population~III stars. This subsequently terminates
the Population III era, hence they contribute $\gtrsim 50\%$ of the
ionizing photons only for $z \gtrsim 30$, and at $z=10$ contribute
$<1\%$.  The Population~III contribution can be increased by delaying
metal mixing into the ISM. However comparing the resulting metal
abundance pattern to existing measurements in $z\lesssim 6$ DLAs, we
show that the fractional contribution of high-mass Population~III
stars to the ionization rate must be $\lesssim 10\%$ at $z=10$. Future
abundance measurements of $z\sim 7$--$8$ QSOs and GRBs should probe
the era when the chemcial vestiges of Population~III star formation
become detectable.
\end{abstract}

\keywords{Cosmology: dark ages, reionization, first stars ---
  galaxies: evolution --- galaxies: ISM --- stars: Population~III}

\section{Introduction}
\label{s:introduction}

Reionization of the intergalactic medium (IGM) is a watershed event in
the history of the universe, and is tightly coupled to the problem of
galaxy formation at high redshift.  The primary evidence for hydrogen
reionization comes from observation of Gunn-Peterson troughs
\citep{1965ApJ...142.1633G} in spectra of high-redshift quasars
\citep{2006AJ....132..117F, 2011Natur.474..616M}.  Other pieces of
evidence are in the angular power spectrum of polarization anisotropy
of the CMB \citep{2011ApJS..192...16L, 2012arXiv1212.5226H}, evolution
of the IGM temperature \citep{2003ApJ...596....9H,
  2011MNRAS.410.1096B, 2012MNRAS.421.1969R}, and evolution of the
luminosity function of Lyman-$\alpha$ emitters
\citep{2010ApJ...723..869O}. For a more complete discussion of
observational constraints on reionization see, e.g., review by
\citet{2006ARA&A..44..415F}.  A general picture that emerges from
these observations and a broad class of theoretical models is that
H~\textsc{i} reionization was a gradual process that lasted for
hundreds of Myr from $z\sim 20$ to $z\sim 6$, and that star-forming
galaxies most likely provided the required ionizing photons
\citep{2006MNRAS.371L..55C, 2011MNRAS.413.1569M, 2012MNRAS.423..862K,
  2013ApJ...768...71R}.

Given this evidence, several recent studies have used averaged
radiative transfer models to ask whether the observed populations of
galaxies at high redshift produce enough ionizing photons to reionize
the IGM \citep{2010Natur.468...49R}.  \citet{2012MNRAS.423..862K}
studied constraints from measurements of the hydrogen photionization
rate, $\Gamma_\mathrm{HI}$, from the (post-reionization)
Lyman-$\alpha$ forest \citep{2008ApJ...688...85F, 2013arXiv1307.2259B}
and the requirement that $\Gamma_\mathrm{HI}$ should evolve in a
continuous manner through the epoch of reionization.  They concluded
these two conditions require either a rapid evolution in the ionizing
photon escape fraction, $f_\mathrm{esc}$, or an extrapolation of the
galaxy luminosity function to extremely faint luminosities
($M_\mathrm{UV}\sim -10$ or $L\sim0.002L_*$).  Several independent
studies have come to similar conclusions \citep{2012MNRAS.425.1413F,
  2012MNRAS.tmpL...1M, 2012MNRAS.427.2464F, 2012MNRAS.421.1969R,
  2012ApJ...746..125H, 2012ApJ...759L..38A, 2013ApJ...768...71R}.
This work suggests that (1) $f_\mathrm{esc}$ is at least ten times
higher at $z=10$ than at $z=4$, and/or (2) there is a large population
of undetected faint galaxies that produces the lion's share of the
total ionizing flux, and/or (3) new galactic sources, such as
mini-quasars or Population~III stars, are active at high redshift and
assist star-forming galaxies in reionizing the IGM.

Owing to their primordial composition, Population~III stars have
harder spectra and thus emit more hydrogen-ionizing photons.  A
cluster of Population~III stars (with 100--260 M$_\odot$ Salpeter IMF)
produces an order of magnitude more hydrogen-ionizing photons than a
cluster of Population~II stars (with 0.1--100 M$_\odot$ Salpeter IMF)
with the same total mass \citep{2002A&A...382...28S}.  Thus, if the
Population~III star formation rate is high enough, their contribution
to the total ionizing photon budget could be significant.  In this
paper, we study the impact of Population~III stars on reionization
using a semi-analytic model of galaxy formation that tracks galactic
chemical evolution and is fully coupled to the evolution of the
thermal and ionization state of the IGM.  Tracking the chemical
evolution is crucial for understanding the contribution of Population
III stars, which in our model form when the gas-phase metallicity of
interstellar media (ISM) of galaxies is below a critical metallicy
$Z_{\rm crit}$.  It is these stars in the very first galaxies that
presumably initiated the process of reionization as they were likely
the first sources of hydrogen-ionizing photons.  As the formation of
these stars depends on the metallicity of the gas out of which they
form, the total constribution of these stars to the cosmic star
formation rate (SFR) density, and hence the ionizing photon budget,
depends on the chemical evolution of their environment, which is
precisely what our model aims to calculate.

The role of Population~III stars in cosmic chemical and ionization
evolution has been studied previously.  Faced with a high value of the
Thomson scattering optical depth to the last scattering surface,
$\tau_e$, reported by the first-year WMAP results $\tau_e=0.17$
\citep{2003ApJS..148..175S}, which suggested that reionization occured
very early, several studies considered the ionizing emissivity of
Population~III stars \citep{2003ApJ...584..621V, 2003ApJ...588L..69W,
  2003ApJ...591L...5C, 2004MNRAS.350...47S, 2004ApJ...605..579Y}, and
others considered reionization scenarios driven by Population~III
stars \citep{2004ApJ...617..693D, 2005MNRAS.361..577C,
  2006ApJ...647..773D, 2009MNRAS.398.1782R, 2011MNRAS.413.1569M}.  Due
to the widely different methods and assumptions of these works, it is
is not straigthforward to compare their results. Nonetheless, there
are some common features in their results.  In all of these models,
reionization is initiated by Population~III stars, which dominate the
cosmic stellar content for some time.  Eventually, however, the
Population~III SFR is reduced due to chemical enrichment of the
star-forming gas, and the process of reionization is completed by
Population~II stars.  A general conclusion of these studies was that,
under conservative assumptions regarding various feedback processes
(chemical, star formation, photoionization), Population~III stars
could contribute significantly to hydrogen reionization at $z\gtrsim
10$.

However, none of these models track the chemical evolution of the
galaxies that host Population~III stars at high
redshift.\footnote{Models of Population~III star formation that
  implement chemical evolution exist in the literature
  \citep[e.g.,][]{2007MNRAS.381..647S}.  These have proven to be very
  useful in studying, e.g., the extremely metal-poor stars in the
  Galactic halo.  But these models do not calculate IGM reionization
  history.}  This is a significant drawback, because the time-scale
over which the ISM of these galaxies is enriched by the first
generation of Population~III stars directly regulates the cosmic
Population~III star formation rate and hence the contribution of
Population~III stars to reionization.  Indeed,
\citet{2012ApJ...745...50W} recently used radiation hydrodynamics
simulations to argue for just this kind of feedback.  These authors
found that for high-mass Population~III stars (Chabrier IMF with
$M_\mathrm{char}= 100$ M$_\odot$), which produce pair-instability
supernovae, just one supernova is enough to enrich the parent halo to
a metallicity of $10^{-3}$ $Z_\odot$ and prevent further
Population~III star formation.  However, these authors did not study
the implications of this chemical feedback on the contribution of
Population~III star-forming haloes to ionizing photon budget and
reionization. This would require simulating star formation and
chemical feedback for a cosmic ensemble of haloes with a wide range of
masses, and coupling these to the IGM via radiative transfer while
considering various observational constraints on reionization.

To summarize, previous work aiming to understand the contribution of
high-redshift faint galaxies or Population~III star-forming galaxies
to reionization \citep[e.g.][]{2003ApJ...584..621V,
  2005MNRAS.361..577C, 2013ApJ...768...71R} did not consider the
detailed physics of galaxy formation, with chemical evolution and
Population~III star formation, while models that included these
effects \citep{2007MNRAS.381..647S, 2012ApJ...745...50W} did not
couple galaxy formation with IGM reionization.  In
\citet{2013ApJ...772...93K} we presented a a semi-analytic model of
galaxy formation that tracks the chemical evolution of galaxies as
well as the thermal and ioinization evolution of the IGM, and used
this model to argue that measurements of relative abundances in
high-redshift Damped Ly$\alpha$ systems can place interesting
constraints on the Population III IMF. In this paper, we use this
model to study the impact of Population III star-formation on
reionization. Our model improves upon previous work on this subject in
three important ways.  First, it accounts for chemical evolution
within a halo in detail, as part of a semi-analytical model of galaxy
formation, taking into account stellar lifetimes, and inflows and
outflows.  This lets us study the Population~III-to-II transition in
haloes of various masses.  Second, it uses halo mass assembly
histories from cosmological simulations and fully couples galaxy
formation to the thermal and ionization evolution of the IGM.  This
also lets us consider halo-mass-dependent effects like photoionization
feedback.  Third, we consider a range of Population~III IMFs that
presumably bracket the true IMF.  A chemical evolution model with
these three features, coupled self-consistently with thermal and
ionization evolution of the IGM, provides a useful framework to study
the contribution of Population~III stars to reionization.

\section{Modeling the Coupled Evolution of Galaxies and the IGM}
\label{sec:model}

We use a model of galaxy formation and IGM evolution described in
\citet{2013ApJ...772...93K}.  Here we highlight the the main features of
the model, and refer the reader to that paper for additional details.

\begin{enumerate}
\item Average mass assembly histories of dark matter haloes are
  obtained from fitting functions calibrated to cosmological
  simulations \citep{2010MNRAS.406.2267F} for a number of
  logarithmically spaced halo masses.  These assembly histories are a
  function of halo mass at $z=0$.
\item Baryonic evolution is implemented for each halo mass value.  In
  simplest terms, this assumes that a halo (1) accretes baryons
  through cosmological accretion, (2) forms stars from any gas
  contained in the halo for sufficiently long duration, and (3) ejects
  baryons via supernova powered outflows.  This lets us calculate
  various properties---such as metallicity and star formation
  rate---as functions of halo mass.  Global averages, such as the
  cosmic SFR density, are calculated by integrating over all halo
  masses.
\item Gas content of a halo is influenced by gas inflow due to cosmic
  accretion, star formation, stellar mass loss, and gas outflow due to
  supernova feedback.  Gas inflow is calculated as being proportional to the
  dark matter acretion rate as given by the mass assembly history.
  Outflow rates are calculated according to the stellar IMF employed, by
  comparing the kinetic energy released by supernovae with the depth
  of the halo potential well.  We also account for stellar lifetimes
  and mass loss from existing stars.
\item We assume that star formation rate, $\psi$, in a halo tracks the
  total amount of cold gas, $M_\mathrm{cool}$, which is determined by
  defining a metalicity dependent cooling radius
  \citep{1999MNRAS.303..188K, 2003MNRAS.339..312S}.
\item Metal content of a halo is calculated by accounting for
  nucleosynthetic yields by integrating over the Population~II and
  Population~III stellar IMFs.  Halo metallicity is diluted by inflows
  from the metal-poor IGM and is enhanced by stellar nucleosynthesis.
  We do not assume instantaneous recycling in this calculation, i.e.,
  we fully account for the delay in the production of metals due to
  finite stellar lifetimes.  In this work, our fiducial model assumes
  instantaneous and homogeneous mixing of metals in the ISM.  But we
  also consider variations of the model which relax this assumption.
\item We implement Population III stars using a critical metallicity
  argument $Z_\mathrm{crit}$, with our fiducial value as
  $Z_\mathrm{crit}=10^{-4} Z_\odot$. When the ISM metallicity in a
  halo becomes larger than $Z_\mathrm{crit}$, Population~III star
  formation stops, and new stars form according to a Population~II
  IMF.
\item We consider two different Population~III IMFs
  \citep{2009Sci...325..601T, 2011ApJ...729L...3D}: 1--100 M$_\odot$
  Salpeter and 100--260 M$_\odot$ Salpeter.  These are selected to
  represent two extreme possibilities.  (We discuss this choice
  below.)  The Population~II IMF is kept constant at 0.1--100
  M$_\odot$ Salpeter.
\item Stellar lifetimes are taken from \citet{1989A&A...210..155M} and
  \citet{2002A&A...382...28S}.  Metal yields are taken from
  \citet{2002ApJ...567..532H} and \citet{1995ApJS..101..181W}.
  Population II SEDs are synthesised using {\scshape starburst99}
  \citep{1999ApJS..123....3L, 2005ApJ...621..695V} with respective
  metallicities.  Synthetic spectra of Population~III stars are taken
  from \citet{2002A&A...382...28S}.
\item We model the thermal, ionization, and chemical evolution of the
  IGM by implementing an inhomogeneous IGM with a lognormal density
  distribution.  We calculate the evolution of volume filling factor
  of H~\textsc{ii} regions, $Q_{\rm HII}(z)$, according to the method
  outlined by \citet{2000ApJ...530....1M}.  Reionization is said to be
  complete when all low-density regions are ionized (we denote the
  corresponding redshift of reionization by $z_\mathrm{reion}$.).  The
  corresponding IGM thermal evolution is determing by solving the
  thermal evolution equation \citep{2000ApJ...534..507C,
    2003ApJ...596....9H}, accounting for photoheating and Compton and
  recombination cooling.  The minimum mass of star-forming regions is
  calculated from the IGM temperature using the Jeans criterion
  \citep{2001PhR...349..125B}.  Chemical evolution of the IGM is
  calculated by accounting for outflows from haloes of all masses.
\end{enumerate}

Before discussing our results, we briefly comment on the
Population~III IMFs considered in this work.  The IMF of
Population~III stars is poorly understood. Several early studies
predicted that Population~III stars have a characteristic mass of a
few hundred M$_\odot$ \citep[e.g.,][]{2002Sci...295...93A}.  In recent
years, this prediction has come down in the range of 20--100 M$_\odot$
\citep[e.g.,][]{2009Sci...325..601T}.  Recently,
\citet{2011ApJ...729L...3D} argued that in the absence of metal-line
cooling, dynamical effects can still lead to fragmentation in
proto-stellar gas clouds.  In their simulations, this resulted in
Population~III stars with masses as low as 0.1 M$_\odot$ (this
picture predicts the existence of many Population~III stars surviving
in the Galaxy till the present day.).  Given our current poor understanding
of the Population~III IMF, in this work we choose to
work with two different IMFs: 1--100 M$_\odot$ Salpeter (``low-mass
IMF'') and 100--260 M$_\odot$ Salpeter (``high-mass IMF'').  The
former has metal yields from AGB stars and core-collapse supernovae,
while the latter has yields from pair-instability supernovae.

\section{Results}
\label{sec:results}

We now present the results of our model.  In Section
\ref{sec:contrib}, we describe the results of our fiducial model which
assumes instantaneous and homogeneous mixing of metals in the ISM, and
show that it predicts a very small Population~III contribution to
reionization.  The main reason for this is chemical feedback, which we
study in Section \ref{sec:sets}.  We then study the conditions under
which Population~III stars could contribute significantly to
reionization, by modelling delayed mixing of metals in the ISM,
thereby reducing chemical feedback.  In Section \ref{sec:chemconst} we
look at the constraints from chemical enrichment measurements of high
redshift DLAs and argue that these rule out significantly delayed
mixing, and hence any significant contribution from high-mass
Population~III stars.

\subsection{Population~III Contribution to Reionization}
\label{sec:contrib}

Our model calibration procedure is described in detail in
\citet{2013ApJ...772...93K}.  The model is calibrated by matching (1)
the observed cosmic SFR density evolution \citep{2006ApJ...651..142H},
(2) the fraction of total baryon density in collapsed haloes at $z=0$
\citep{2004ApJ...616..643F}, and the reionization history of the IGM
as measured by (3) the electron Thomson scattering optical depth to
the last scattering surface \citep{2012arXiv1212.5226H} and (4) the
hydrogen photoionization rate evolution \citep{2004MNRAS.350.1107M,
  2007MNRAS.382..325B, 2008ApJ...688...85F, 2013arXiv1307.2259B}.

We assume a constant star-formation efficiency parameter $f_*$,
defined by
\begin{equation}
  \psi = f_*\left(\frac{M_\mathrm{cool}}{t_\mathrm{dyn}}\right),
  \label{eqn:sf}
\end{equation}
where $\psi$ is the halo star formation rate, $M_\mathrm{cool}$ is the
cold gas mass in the halo, and $t_\mathrm{dyn}$ is the halo dynamical
time.  The star-formation efficiency is then tuned to reproduce the
cosmic SFR density.  For our fiducial model (described below), we have
$f_*$ (Pop.~II) $=0.002$ and $f_*$ (Pop.~III) $=0.004$.  Star
formation in haloes is accompanied by supernova-induced outflows, the
strength of which is fixed by matching the fraction of total baryon
density in collapsed haloes at $z=0$.

Once the star formation efficiency is fixed, reionization depends
mainly on the escape fraction of ionizing photons, $f_\mathrm{esc}$,
defined as the fraction of ionizing photons that escape the parent
galaxy.  We assume $f_\mathrm{esc}$ to be independent of redshift and
halo mass, and calibrate it to reproduce the observed values of the
electron Thomson scattering optical depth, $\tau_e$ and the hydrogen
photoionization rate, $\Gamma_\mathrm{HI}$.  It is crucial to include
both these constraints as the observed hydrogen photoionization rate
evolution imposes continuity on the epoch of reionization.  Our
fiducial model has $f_\mathrm{esc}=0.2$.  Although, in our calibrated
model, the IGM electron Thomson scattering optical depth $\tau_e$ is
always obtained in the 1$\sigma$ range of its best-fit value
($0.089\pm 0.014$ as given by WMAP9; \citealt{2012arXiv1212.5226H}),
our value of $\tau$ is always closer to the upper-limit of this range.
This suggests that an evolving escape fraction is necessary for
fitting both $\tau_e$ and $\Gamma_\mathrm{HI}$ simultaneously
\citep{2012MNRAS.tmpL...1M, 2012MNRAS.427.2464F}.

We calibrate the model independently for each of the two
Population~III stellar IMFs considered in this paper.  Our fiducial
model assumes instantaneous mixing of metals in the ISM.  We describe
the results of this model in this section.  We discuss the effect of
relaxing the instantaneous mixing assumption in subsequent sections.

\begin{figure*}
  \begin{center}
    \includegraphics*[width=\textwidth]{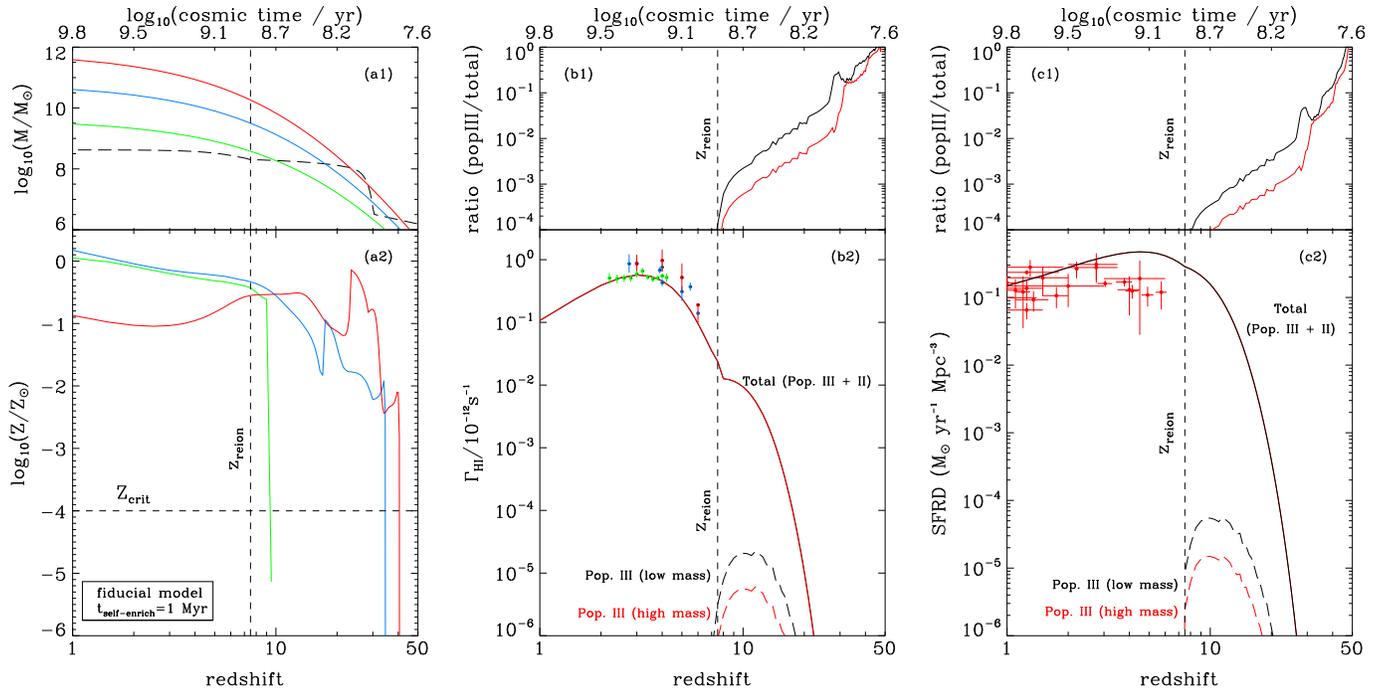}
  \end{center}
  \caption{Results for our fiducial model. Panels (a1) and (a2) show
    the metallicity evolution for three particular halo masses for
    illustration.  The long-dashed curve in panel (a1) shows the
    evolution of $M_\mathrm{min}$ in cosmological H~\textsc{ii}
    regions.  Panel (b1) and (b2) show the evolution of the hydrogen
    photoionization rate and the contribution to it from
    Population~III stars for the two models considered.  The
    corresponding panel (b1) plots the ratio of the two quantities.
    In this panel, the black curve corresponds to the low-mass IMF
    model and the red curve to the high-mass IMF model.  Panels (c1)
    and (c2) show the evolution of the cosmic SFR density in the model
    and the contribution from Population~III stars.  Vertical dashed
    lines in all panels show the redshift of reionization.}
  \label{fig:fiducial}
\end{figure*}

\subsubsection{Population~III SFR}

Fig.~\ref{fig:fiducial} shows the results of our fiducial model.
Panel (c2) shows the predicted total (Population~II $+$
Population~III) cosmic SFR density evolution for the two different
Population~III IMFs (low-mass IMF in black and high-mass IMF in red).
The red data points are observational measurements from a compilation
by \citet{2006ApJ...651..142H}, in which consistent dust obscuration
corrections, SFR calibrations, and IMF assumptions are applied to
ultraviolet and far-infrared data.  Our model predictions are in good
agreement with the data regardless of what Population~III IMF is
assumed.  Indeed, the total SFR density in the two cases is nearly
identical (the red and black solid curves overlap), because the
contribution of Population~III star formation to the total SFR is very
low over most of the cosmic history.  The Population~III SFR density
evolution is shown separately by the dashed curves.  The corresponding
panel (c1) shows the fraction of the total SFR produced by
Population~III star formation.  We find that the contribution of
Population~III stars to the total SFR density is low (less than 10\%)
throughout most of the age of the Universe.  It is greater than 10\%
for less than 100 Myr ($z>40$).  Population~III stars dominate the
cosmic SFR density for an even smaller time period at $z\sim 50$
\citep[cf.][]{2006MNRAS.373L..98N}.  By the time that galaxies are
observable at $z<8$, all haloes capable of forming stars (i.e., above the
minimum mass for star-formation) have already enriched themselves
above the critical metallicity, and hence the Population~III SFR is
zero at these late times. Thus, although Population~III stars initiate
the epoch of reionization, they quickly relinquish their dominant role
by enriching their environment.  We discuss the reason behind this
below.

We note that the star formation history shown in
Fig.~\ref{fig:fiducial}, together with our chemical evolution model,
is also consistent with the observed mass-metallicity relations
\citep{2006ApJ...644..813E, 2008A&A...488..463M} at redshifts 2.3 and
3.7 (see \citealt{2013ApJ...772...93K} for details).  The
corresponding IGM metallicity is also consistent with observational
estimates \citep{2003ApJ...596..768S, 2004ApJ...606...92S}.  We also
note here that the difference between the two Population~III SFR
density curves (dashed curves) in panel (c2) of
Fig.~\ref{fig:fiducial} can be understood from the different metal
yields for the two Population~III IMFs (recall that the Population~II
IMF is the same in both cases).  The high-mass IMF has a larger metal
yield.  As a result, in this case haloes are enriched beyond
$Z_\mathrm{crit}$ earlier as compared to the low-mass IMF, and
Population~III star formation is terminated earlier.  This is
reflected in the reduced level of Population~III SFR density for the
high-mass IMF.

\subsubsection{Photoionization rate}

The small contribution of Population~III stars to the total cosmic SFR
density suggests that their contribution to reionization will also be
small.  This is seen in panel (b2) of Fig.~\ref{fig:fiducial}, which
shows the evolution of the hydrogen photoionization rate in our model.
Making the so-called local source approximation, which is valid for
spectral indices typical to star-forming galaxies, the hydrogen
photoionization rate is given by
\begin{equation}
  \Gamma_\mathrm{HI}(z)=(1+z)^3\int_{\nu_{912}}^\infty d\nu \lambda(z,\nu)
  \dot n_\nu(z) \sigma(\nu),
\end{equation}
where $\sigma(\nu)$ is the photoionization cross-section of hydrogen,
$\lambda(z,\nu)$ is the redshift-dependent mean free path of ionizing
photons.  We calculate the mean free path of ionizing photons as in
Paper 1 by integrating over the lognormal density PDF of the IGM and
estimating the average distance between high-density, neutral regions.
This method is calibrated to reproduce the incidence rate of
Lyman-limit systems at low redshifts \citep{2000ApJ...530....1M,
  2005MNRAS.361..577C, 2009ApJ...705L.113P}.  Our model agrees with
the measurement of \citet{2009ApJ...705L.113P}.  The quantity $\dot
n_\nu(z)$ is the number density of ionizing photons in the IGM per
unit time, and is given by
\begin{equation}
  \dot n_\nu(z) = f_\mathrm{esc}\dot\rho_*(z)\int
  dm\,\phi(m)t_\mathrm{age}(m)Q_H(m).
\label{eqn:flux}
\end{equation}
The integral in this equation is over stellar masses and takes the
IMF-dependence of the ionizing photon flux into account: $\phi(m)$ is
the normalised stellar IMF, $t_\mathrm{age}(m)$ is the age of a star
with mass $m$ and $Q_H(m)$ the stellar hydrogen-ionizing photon flux
(in photons s$^{-1}$) provided by stellar evolution models
\citep{2002A&A...382...28S}.  The quantity $f_\mathrm{esc}$ is the
escape fraction of ionizing photons that accounts for the fraction of
the ionizing photons that escape into the IGM.  The escape fraction is
a free parameter of our model.  In this paper, we assume that
$f_\mathrm{esc}$ is constant at all redshifts.  We comment on the
effect of this assumption below.  Our fiducial model has
$f_\mathrm{esc}=0.2$

In panel (b2) of Fig.~\ref{fig:fiducial}, the data points are
measurements of the hydrogen photoionization rate as deduced from the
mean opacity of the hydrogen Ly$\alpha$ forest
\citep{2004MNRAS.350.1107M, 2007MNRAS.382..325B, 2008ApJ...688...85F},
where we note that there is disagreement between the measurements at
the factor-of-two level, which likely results from different
assumptions about the density distribution and thermal state of the
IGM. Here were choose to fit the data by \citet{2008ApJ...688...85F}
but this choice is not critically important for our main result.  The
solid curves in panel (b2) of Fig.~\ref{fig:fiducial} show the model
predictions corresponding to two different Population~III IMFs
(low-mass IMF in black and high-mass IMF in red).  The model
predictions agree very well with the observational
measurements.\footnote{Note that we have ignored the ionizing photon
  contribution from quasars in our model, which are expected to only
  contribute significantly for $z<3$ \citep{2012MNRAS.425.1413F} and
  we do not expect any Population III contribution at these late
  times.}  As seen by the evolution of the hydrogen photoionization
rate, reionization in our model is gradual.  It begins at $z\gtrsim
30$ and is $90\%$ complete by $z\sim 7$.  This gradual change in the
ionization state of the IGM helps us simultaneously reproduce the
observed electron scattering optical depth $\tau_e=0.089\pm 0.014$
\citep{2012arXiv1212.5226H} and the photoionization rate data.  Before
reionization at $z_\mathrm{reion}=7.5$, the photoionization rate
increases rapidly as UV photon sources build up.  There is a sudden
jump at $z=z_\mathrm{reion}$ when different H~\textsc{ii} regions
overlap.  (This redshift is marked by the vertical dashed line in
Fig.~\ref{fig:fiducial}).  This is because at this redshift, a given
point in the IGM starts ``seeing'' multiple sources, which rapidly
enhances the UV photon mean free path, thereby affecting the
photoionization rate.  Secondly, Fig.~\ref{fig:fiducial} also shows
that the contribution of Population~III stars to reionization is
small.  This is clear from the overlap between the red and black solid
curves in panel (b2), and is evident in the dashed curves, which show
the contribution to $\Gamma_\mathrm{HI}$ by Population~III stars.  We
see that the Population~III contribution to photoionization rate is
subdominant over most of the reionization history.  Panel (b1) of
Fig.~\ref{fig:fiducial} further highlights this by showing the
fraction of the H~\textsc{i} photoionization rate contributed by
Population~III star-formation, relative to the total rate.  Except at
the earliest stages of galaxy formation ($z\sim 50$), the ratio is
much less than unity.  For the low-mass IMF, the ratio is less than
10\% for $z\lesssim 20$, and for the high-mass IMF, it is less than
10\% for $z\lesssim 30$.

\subsubsection{Understanding the Low Population~III Contribution}

We now discuss the reason behind the small contribution of
Population~III stars to reionization in our model.  This contribution
depends on three parameters of the model: (1) the efficiency with
which cold gas is converted into stars, (2) the escape fraction of
ionizing radiation, and (3) chemical feedback, quantified by the time
scale over which a halo enriches itself beyond the critical
metallicity $Z_\mathrm{crit}$ and stops the formation of
Population~III stars.  These three factors are not independent of each
other: a higher efficiency of star formation reduces the
self-enrichment time scale of a halo, and a higher $f_\mathrm{esc}$
requires us to reduce the star formation efficiency if we are to
satisfy the observational constraints on reionization.  In our model,
the constraints from $\Gamma_\mathrm{HI}$, $\tau_e$ and the cosmic SFR
density completely fix the star formation efficiency and the escape
fraction.  Therefore, as we assume instantaneous metal mixing in our
fiducial model, the self-enrichment time scale is also fixed.

Panel (a2) of Fig.~\ref{fig:fiducial} shows the metallicity evolution
of three different haloes in our model, which helps one understand the
small contribution of Population~III stars to reionization in our
model (for simplicity, we show results only for the high-mass IMF in
this panel, but the low-mass IMF gives similar trajectories). The
horizontal dashed line in this panel shows the critical metallicity
$Z_\mathrm{crit}$, which we always take to be $10^{-4} Z_\odot$.  The
mass assembly history of these haloes is shown in the corresponding
panel (a1); their masses at $z=0$ are about $10^{10}$ M$_\odot$,
$10^{11}$ M$_\odot$, and $10^{12}$ M$_\odot$ respectively.  The dashed
curve in the panel (a1) shows the evolution of the minimum mass of
star-forming haloes, $M_\mathrm{min}$, which is set according to the
Jeans criterion, which in turn depends on the thermal evolution of the
IGM \citep[see, e.g.,][for a recent discussion]{2013arXiv1305.0210R}.
To be specific, here we consider haloes that collapse in cosmological
H~\textsc{ii} regions, and the $M_\mathrm{min}$ evolution shown
corresponds only to these regions (but the discussion could be easily
generalized to H~\textsc{i} regions).  In these regions,
$M_\mathrm{min}$ is roughly constant at $\sim 10^{8}$ M$_\odot$ as
this evolution is only determined by the Jeans scale corresponding to
the characteristic temperature of the IGM at $T\sim 10^4$ K.  In each
halo's assembly history, star formation begins when its mass crosses
the threshold set by $M_\mathrm{min}$.  This is manifest in the
metallicity evolution history of each halo as shown in panel (a2).  It
is seen that the halo metallicities increase mostly
monotonically.\footnote{Some non-monotonicity seen in panel (a2) of
  Fig.~\ref{fig:fiducial} is due to the behaviour of $M_\mathrm{min}$
  at the highest redshift, which can restrict gas inflow into haloes
  at certain times, affecting the metallicity.}  Additionally, in each
case, the initial burst of Population~III stars is sufficient to
rapidly enhance the ISM metallicity beyond $Z_\mathrm{crit}$.  This
metallicity evolution is in good agreement with the hydrodynamic
simulations of \citet{2012ApJ...745...50W}.  It is exactly this prompt
enrichment of the ISM of high-redshift galaxies which causes an early
cut-off in the Population~III SFR.  The range of redshifts in panel
(c2) that show sub-dominant Population~III SFR, arises from only the
lowest mass star-forming halos, which only recently crossed the
minimum mass threshold. However, the vast majority of star-forming
halos at higher masses have already been chemically self-enriched by
this time, and have stopped forming Population~III stars, which in
turn reduces the overall contribution of Population~III stars to the
photoionization rate.  As previously mentioned above, apart from the
star formation efficiency assumed in the model, the rapid enrichment
of galaxies is a result of the assumption that metals are
instantaneously mixed in the ISM gas.  We discuss the assumption in
detail in the next section.

Finally, we briefly note that haloes down to a mass limit of
$M_\mathrm{min}\sim 10^8$ M$_\odot$ contribute to the ionization flux
in Eqn. (\ref{eqn:flux}).  At $z\sim 7$ this corresponds to a UV
magnitude to $M_{1500}=-10$ in our model, which agrees well with the
very faint minimum galaxy magnitude, to which the UV luminosity
function of Lyman break galaxies (LBGs) must be extrapolated to in
order to reionize the universe \citep{2012MNRAS.423..862K,
  2012MNRAS.425.1413F, 2013ApJ...768...71R}. Furthermore, the
halo masses which we deduce for these faint galaxies 
$M_\mathrm{min}\sim 10^8$ M$_\odot$ agree well with masses
deduced  using abundance matching
\citep{2010ApJ...714L.202T, 2012MNRAS.423..862K}.  
This indicates that in our fiducial model, it is the faint galaxies
that produce the bulk of the ionizing photons that reionized the IGM, 
and not Population~III stars. 

\subsection{The Halo Self-Enrichment Timescale}
\label{sec:sets}

\begin{figure}
  \begin{center}
    \includegraphics*[width=\columnwidth]{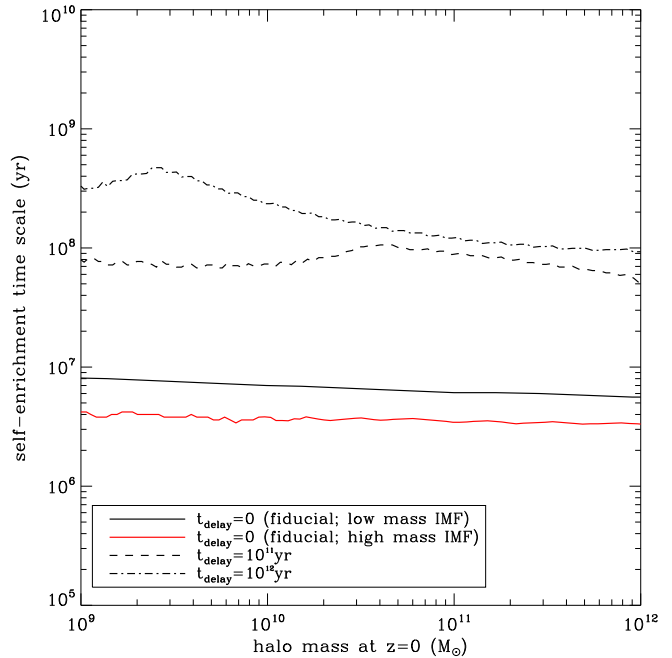}
  \end{center}
  \caption{Halo self-enrichment time scale in our model, defined as
    the time between the first star-formation episode of the halo and
    the time at which its gas-phase metallicity crosses
    $Z_\mathrm{crit}$.  It quantifies the time scale of Population~III
    star formation in a given halo. Halo mass at $z=0$ serves as a
    label for different haloes.  The solid curves show the time scale
    for the fiducial models.  Other curves show the time scale for
    models with delayed enrichment.}
  \label{fig:ztime}
\end{figure}

\begin{figure*}
  \begin{center}
    \includegraphics*[width=\textwidth]{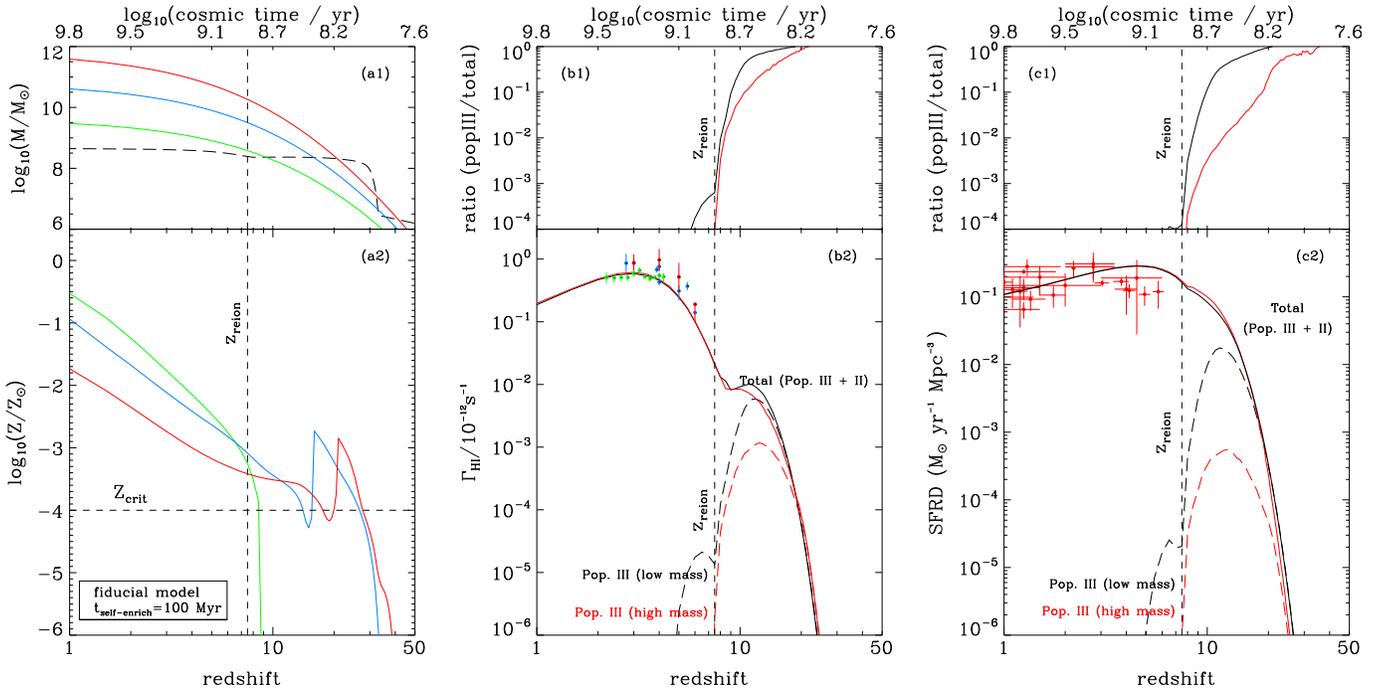}
  \end{center}
  \caption{Same as Fig.~\ref{fig:fiducial} but for a model in which
    metal enrichment in ISM is gradual with a delay time of
    $t_\mathrm{delay}=10^{11}$ yr.  This increases the self-enrichment
    time scale of haloes to about $10^8$ yr.  Consequently, the
    contribution of Population~III stars to the cosmic star formation
    rate and to the hydrogen photoionization rate is enhanced
    relatived to the fiducial model. See text for details.}
    \label{fig:t8}
\end{figure*}

\begin{figure*}
  \begin{center}
    \includegraphics*[width=\textwidth]{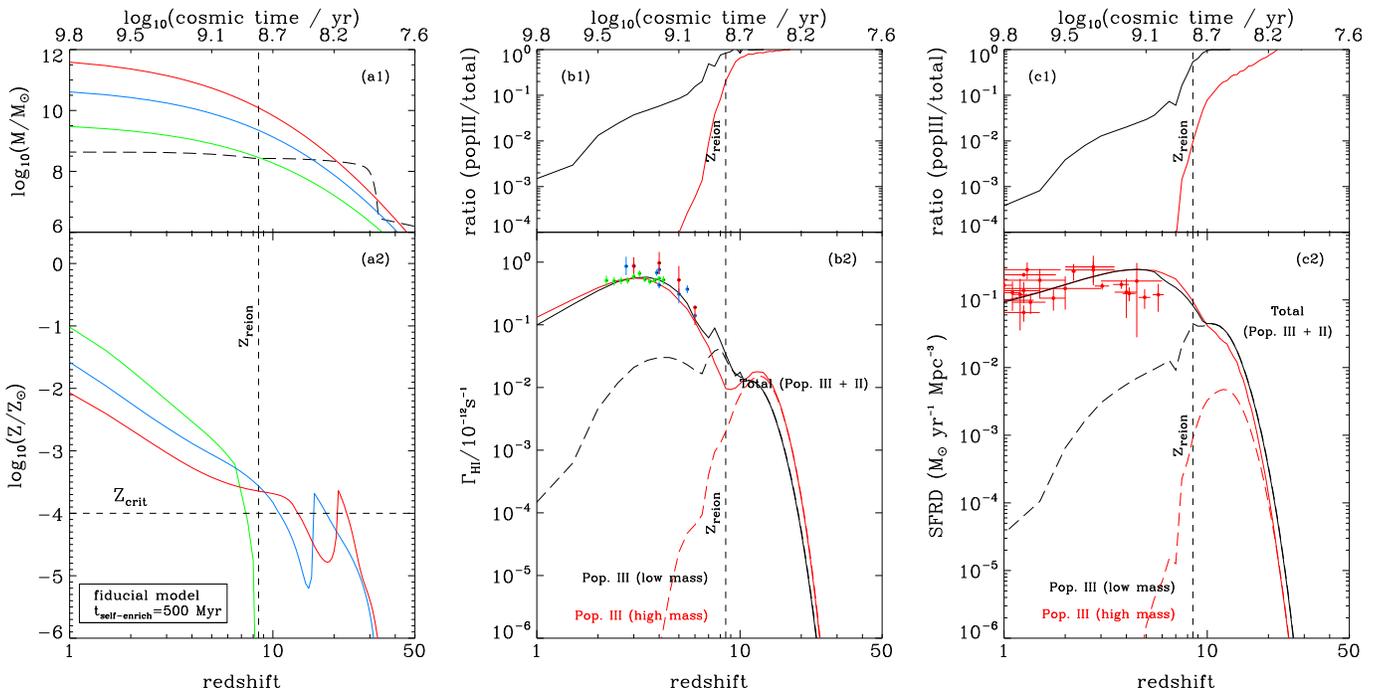}
  \end{center}
  \caption{Same as Figs.~\ref{fig:fiducial} and \ref{fig:t8} and but
    for a model in which metal enrichment in ISM is gradual with a
    delay time of $t_\mathrm{delay}=10^{12}$ yr, i.e., an
    order-of-magnitude slower that in Fig.~\ref{fig:t8}.  This
    further increases the self-enrichment time scale of haloes to
    about $4\times 10^8$ yr.  As a result, the contribution of
    Population~III stars to the cosmic star formation rate and to the
    hydrogen photoionization rate is high enough to complete hydrogen
    reionization.  Note that in this model, the redshift of
    reionization is slightly different for the two Population~III
    IMFs; we only show the low-mass IMF case for simplicity.}
  \label{fig:t9}
\end{figure*}

In the results presented above, metals injected into the ISM by
supernova explosions are assumed to be instantaneously mixed into the
halo ISM.  Instantaneous and homogeneous mixing of metals is a
standard assumption in galactic chemical evolution studies
\citep{1980FCPh....5..287T, 2009nceg.book.....P, 2012ceg..book.....M}
and is responsible for the early termination of Population~III star
formation in our fiducial model.  Note that we are not assuming
instantaneous recycling, since we directly model the delay in the
synthesis of new metals due to finite stellar lifetimes. However, we
have assumed instantaneous mixing, which is to say that after a
supernova event, newly liberated metals are instantneously available
in the ISM to influence the next generation of star-formation.  We
have seen in our fiducial model, that the initial burst of
Population~III stars is sufficient to enhance the ISM metallicity
beyond $Z_\mathrm{crit}$ in any halo mass bin over a very short time
scale.  When this happens, the corresponding galaxy stops forming
Population~III stars.  Since most of the mass in the universe is
contained in $M_*$ haloes, the Population~III cosmic SFR density
begins to decline as soon as these haloes cross the $Z_\mathrm{crit}$
threshold.  This is the crucial effect that reduces the contribution
of Population~III stars, and that was not captured in previous models
because they did not track chemical enrichment for a large population
of haloes in a cosmological volume.  We now discuss the dependence of
this result on our assumption of instantaneous mixing.

The solid lines in Fig.~\ref{fig:ztime} show the self-enrichment time
scale as a function of halo mass for the two Population~III IMFs in
our fiducial model, where the low-mass IMF is shown in black and
high-mass IMF in red.  We define the self-enrichment time scale as the
time between the first star-formation episode of the halo and the time
at which its gas-phase metallicity crosses $Z_\mathrm{crit}$.  In this
plot, haloes are labelled by their mass at $z=0$; their actual mass
value at the time of crossing $M_\mathrm{min}$ is not shown.  In the
low-mass IMF case (solid black), the time scale is a few times $10^6$
yr. Pair-instability SNe have higher metal output, which results in a
shorter enrichment time scale by a factor of 3 in the high-mass IMF
case (solid gray).  As discussed in section \ref{sec:contrib}, it is
this short self-enrichment time scale that restricts the contribution
of Population~III stars to reionization.  Note that these curves have
a very flat dependence on halo mass because the self-enrichment time
scale depends primarily on stellar lifetimes.

In the ISM of a galaxy, mixing of metals ejected by supernovae is
carried out by different mechanisms on different length scales:
diffusive dispersion due to large-scale motion on kpc scales,
turbulent diffusion on pc scales, and molecular diffusion on smaller
scales.  As a result, the self-enrichment time scale is decided by (1)
the respective time scales of these mixing processes, (2) the rate of
metal production by supernovae, and (3) rate of interaction with the
environment, via inflows and outflows of gas.  In our fiducial model,
the time scale of the mixing processes is assumed to be zero and
therefore the self-enrichment time scale depends only on the stellar
lifetimes via the supernova rate.  This is the reason behind the short
self-enrichment time scale shown in Fig.~\ref{fig:ztime}.  We now
relax this instantaneous mixing assumption.

Supernova-driven metal-mixing in the interstellar medium has been
studied using hydrodynamical simulations by
\citet{2002ApJ...581.1047D}, who used tracer particles in a simulation
of a $1\times 1\times 20$ kpc$^3$ region of the Galactic disk and
found mixing time scales between $10^6$ and $10^8$ yr (see their
Fig.~8) for supernova rates of $\sim 1$ to $\sim 100$ times the
Galactic supernova rate.  The lower end of this range is of the same
order of magnitude as the self-enrichment time scale in our fiducial
model.  But the results of these simulations suggest that the mixing
time scale could very well be two or three orders of magnitude larger.

We consider the effect of longer self-enrichment time scales on our
result by implementing a ``mixing function'', $f(t)$, which governs
the rate at which metals are mixed in the ISM.  The source term due to
star formation in the metallicity evolution equation is of the form
(Eqn.~19 in \citealt{2013ApJ...772...93K})
\begin{multline}
   \dot M_{Z}(z) = \int_{m_l}^{m_u}dm\,\phi(m)\cdot\psi[t(z)-\tau(m)]\\\times mp_Z(m),
\label{eqn:sfrz}
\end{multline}
where $m$ is the stellar mass, $\psi(t)$ is the star formation rate at
time $t$, $\phi(m)$ is the stellar IMF, $\tau(m)$ is the stellar age,
and $p_Z(m)$ is the mass fraction of a star of initial mass $m$ that
is converted to metals and ejected.  The limits $m_l$ and $m_u$ define
the range of stellar masses considered in the IMF.  To relax the
instantaneous mixing assumption, we modify Eqn.~(\ref{eqn:sfrz}) to
\begin{multline}
  \dot M_{Z}(z) = \int_{m_l}^{m_u}dm\,\phi(m)mp_Z(m)\\\times\int_{t(z)-t_\mathrm{delay}}^{t(z)}\!\!\!d\tilde t\,\psi[\tilde t(z)-\tau(m)] f(\tilde t),
\label{eqn:sfrz_delayed}
\end{multline}
where $f(\tilde t)$ is a mixing function that serves to delay the
mixing of metals in the ISM after a supernova event.  The mixing
function is defined over a time duration of $t_\mathrm{delay}$.  In
this picture, after a supernova has exploded, the resulting metal mass
is added to the ISM gradually over a time $t_\mathrm{delay}$.  Until
this time, we imagine the metals to be locked up into un-mixed pockets
of the ISM, and so they are not available for future star-formation.
Thus, our fiducial model corresponds to the case where the mixing
function is a delta function with $t_\mathrm{delay}=0$ yr.  For
non-zero values of $t_\mathrm{delay}$, the ISM metallicity can remain
below the critical metallicity $Z_{\rm crit}$ for a longer time
compared to our fiducial model.  Therefore, we would expect that in
this case, Population~III star formation will continue for a longer
period, and possibly impact reionization.  The mixing function is
determined by the complex interplay of various mixing processes in the
ISM.  The most conservative form of the mixing function is a constant
of the form
\begin{equation}
f(t)  =
\left\{
	\begin{array}{ll}
		t_\mathrm{delay}^{-1}  & \mbox{if } t < t_\mathrm{delay} \\
		0 & \mbox{otherwise},
	\end{array}
\right.
\label{eqn:mixfn}
\end{equation}
where $t$ is the time since a supernova explosion.\footnote{This form
  of the mixing function is the most conservative in the sense that
  all time instances from $0$ to $t_\mathrm{delay}$ are treated
  equally.  As a result, metals are mixed into the ISM at a constant
  rate and, in the limit of an exactly closed system, the metallicity
  would increase linearly.}  Note that $f(t)$ is normalised such that
all of the supernova ejecta is mixed in the ISM over the period
$t_\mathrm{delay}$.  We adopt this simple form for the mixing function
in this paper.  Figs.~\ref{fig:t8} and \ref{fig:t9} show the results
of our model with this mixing function and $t_\mathrm{delay}=10^{11}$
yr and $t_\mathrm{delay}=10^{12}$ yr respectively.  The
self-enrichment time scales in these models is compared to that in our
fiducial model in Fig.~\ref{fig:ztime} (only the high-mass
Population~III IMF case is shown for simplicity).  Given that we have
adopted a mixing timescale which is much longer than the age of the
Universe, i.e., $t_\mathrm{delay} = 10^{10}$ or $10^{11}$ yr, the
effective result is that at $z\sim 10$, where $t_{z=10}\sim 10^9$ yr,
these mixing models imply that only 1\% and 0.1\% of the metals
produced by Population~III SNe are mixed into the ISM, respectively.
As we will see below even this small amount of metals is still enough
to produce significant chemical feedback that influences
Population~III star formation.  This results because as shown in panel
(a2) of Fig.~\ref{fig:fiducial}, a single generation of Population~III
SNe inejcts enough metals to raise the the ISM metallicity of $M_*$
halos to a $Z\sim 10^{-1}Z_\odot$ by $z\sim 10$, if these metals are
instantateously mixed. Thus reducing this yield by a factor of $\sim
t_{z=10}/t_\mathrm{delay} \sim 100$, as in the case when
$t_\mathrm{delay} = 10^{11}$, is still sufficient to be in excess of
the critical metallicity. For this reason, and as is also clear from
the mathematical form of Eqn.~\ref{eqn:sfrz_delayed}, we generally
expect a non-linear dependence of the self-enrichment timescale on
$t_\mathrm{delay}$, and similarly the dependence of Population~III
star-formation on $t_\mathrm{delay}$ will also be nonlinear.  For
instance, between Fig.~\ref{fig:t8} and Fig.~\ref{fig:t9}, changing
$t_\mathrm{delay}$ by a factor of 10 results in small changes of
factor of 3 in the self-enrichment time scale.

In Fig.~\ref{fig:t8} the same curves illustrating the star-formation,
reionization, and enrichment history as in Fig.~\ref{fig:fiducial}
are shown but for a model with $t_\mathrm{delay}=10^{11}$ yr.  As
expected, and illustrated in panel (a2) of Fig.~\ref{fig:t8},
haloes now take longer to cross the $Z_\mathrm{crit}$ metallicity
threshold, compared to the fiducial model (panel a2 of
Fig.~\ref{fig:fiducial}, which is on the same scale).  This increases
the cosmic Population~III SFR density, which, although still
subdominant, contributes more than 10\% of the total SFR density at
redshifts above $z\sim 20$ in the high-mass Population~III IMF run
(the corresponding redshift value for the fiducial model was $z\sim
30$).  The relative increase in Population~III star-formation also
manifests as an increase in the hydrogen photoionization rate, which
is also enhanced relative to the fiducial model. Indeed, the
contribution of Population~III stars to the hydrogen photoionization
rate at $z=10$ is 10\% for the high-mass case, whereas 
the contribution was less than 1\% at this redshift in our fiducial
model.  Thus for
$t_\mathrm{delay}=10^{11}$ yr, the contribution of Population~III
stars to reionization is subdominant but significant.  Note that these
numbers are slightly different for the low-mass case for the same
mixing function and $t_\mathrm{delay}$.  In general, low mass
Population~III star formation lasts longer than for the high-mass IMF.
This is because the metal yields of high mass Population~III stars is
higher than those of low mass stars, resulting in larger chemical
feedback, and earlier termination of Population~III. 

Fig~\ref{fig:t9} shows the results of the model with
$t_\mathrm{delay}=10^{12}$ yr, for which the mixing of metal metals is
even more gradual. The impact of Population~III on cosmic
star-formation and the ionizing photon budget is now increased in
magnitude relative to the previous case shown in Fig.~\ref{fig:t8}.
For instance, all star formation for $z\gtrsim 10$ in this model is
Population~III, regardless of the Population~III IMF.  Thus the
process of hydrogen reionization is almost single-handedly carried out
by Population~III stars.  The contribution of Population~III stars to
the hydrogen photoionization rate at $z=10$ is about 60\% for the
high-mass case, and 100\% for the low-mass case.  Their contribution
to the cosmic SFR density at $z=10$ is about 20\% for the high-mass
case, and 80\% for the low-mass case.  Similar to the results in
Fig.~\ref{fig:t8}, the Population~III contribution is higher for
low-mass Population~III IMF because of weaker chemical feedback.

Fig.~\ref{fig:ztime} helps in visualising the effect of the mixing
function on our model.  The self-enrichment time scale in the two
variant models is longer than that in the fiducial model by more than
an order of magnitude.  The increase, however, is still less that
$t_\mathrm{delay}$, as only a fraction of metals are required to
increase halo metallicity beyond $Z_\mathrm{crit}$.

A general lesson from above is that due to constraints imposed by the
measurements of cosmic SFR density, the hydrogen photoionization rate,
and the electron scattering optical depth, the contribution of
Population~III stars to reionization can be enhanced only by
increasing the metal mixing time scale assumed in the model.  This
contribution is generally predicted to be small unless the
self-enrichment time scale is $\gtrsim 3\times 10^8$ yr.

\subsection{Chemical Enrichment Constraints on the Contribution of Population III to Reioinization}
\label{sec:chemconst}

\begin{figure*}
  \plottwo{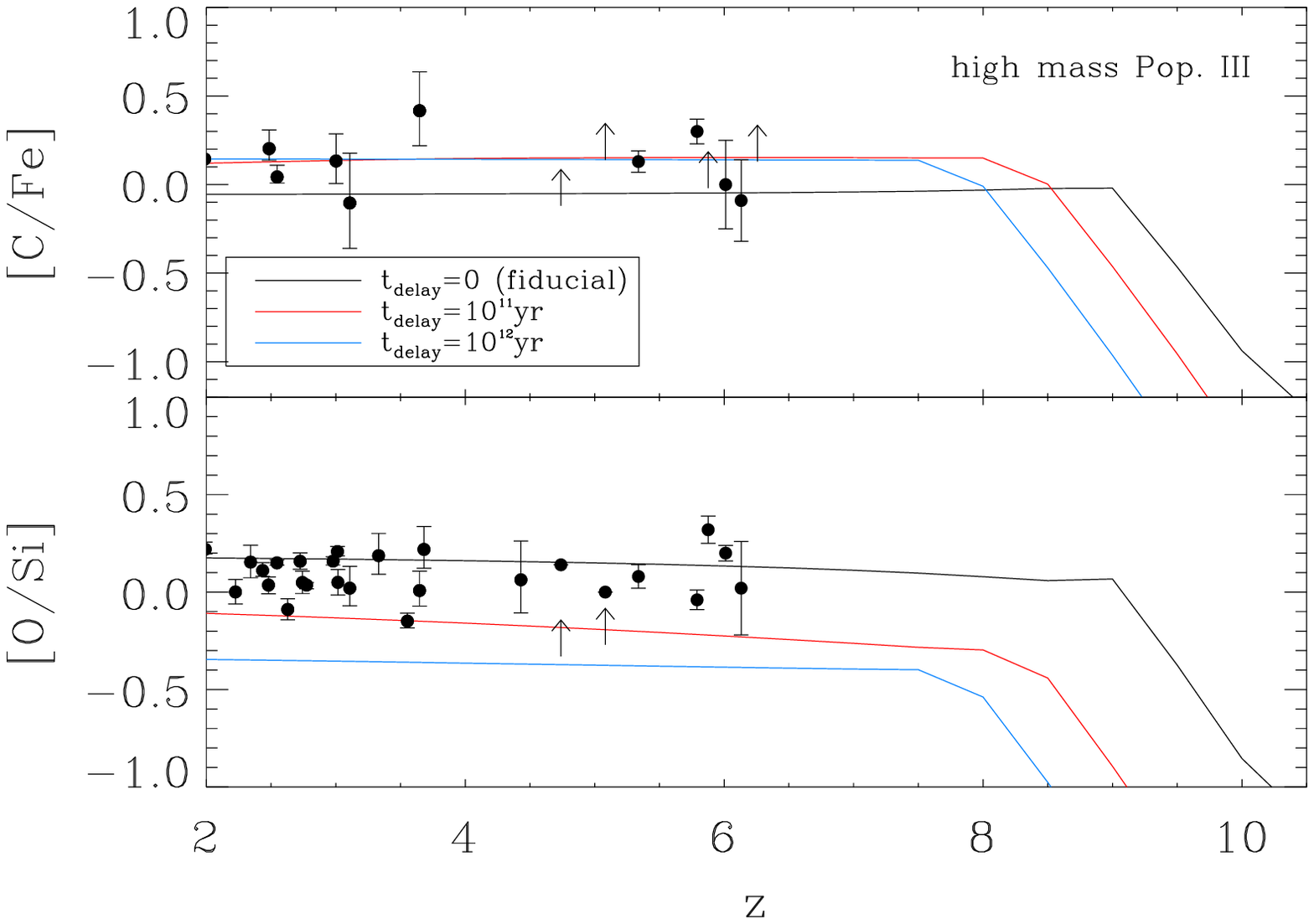}{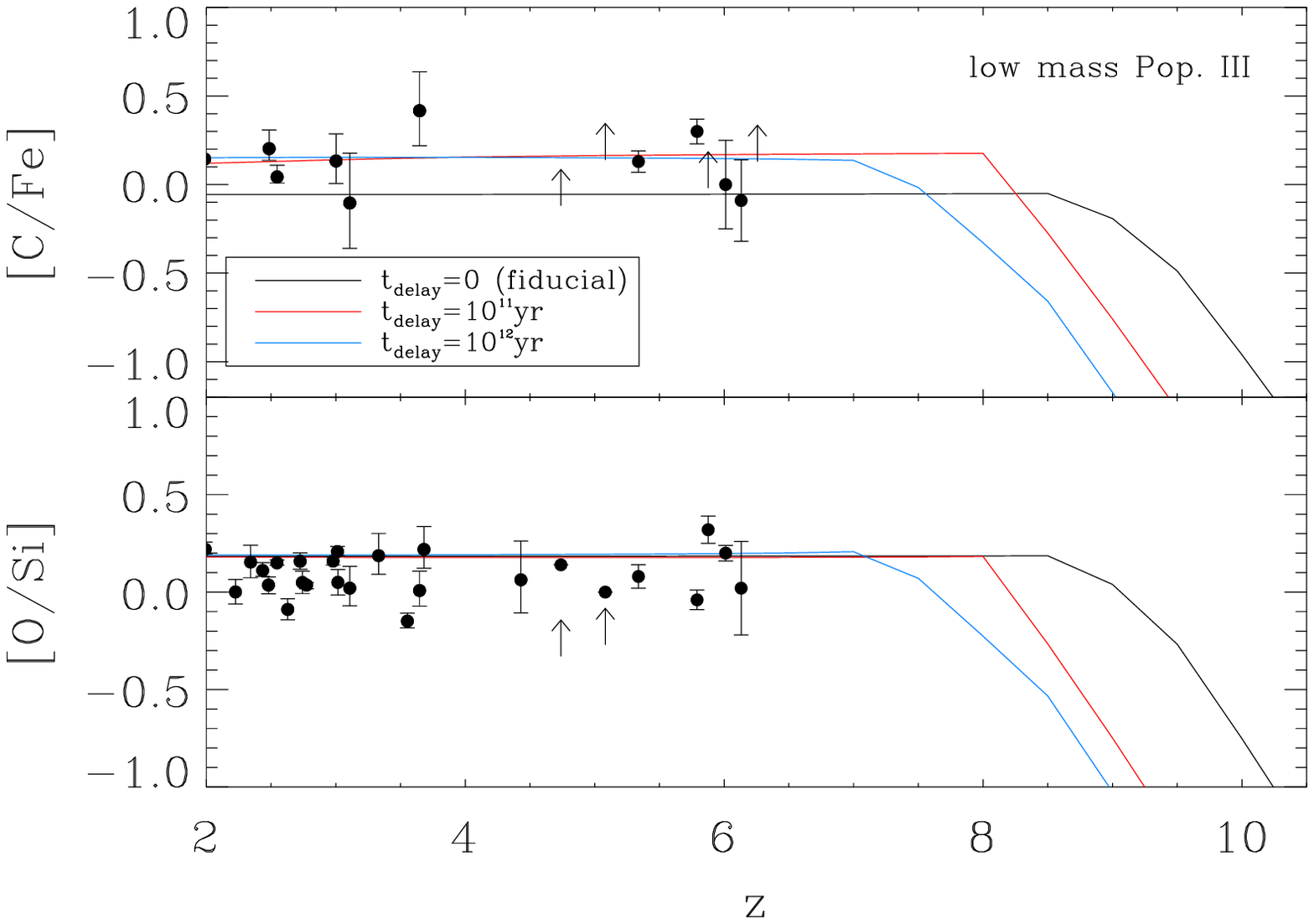}
  \caption{Evolution of mean values of [C/Fe] and [O/Si] DLA relative
    abundances in all models considered in this paper.  Data points
    show observed DLA relative abundances from a compilation by
    \citet{2012ApJ...744...91B}.  Right panel shows low-mass IMF
    models, left panel high-mass IMF models.  Our fiducial model
    (black curves) is consistent with the observational measurements.
    The high-mass IMF model with gradual enrichment with
    $t_\mathrm{delay}=10^{11}$ yr is only marginally consistent with
    the [O/Si] data, while that with $t_\mathrm{delay}=10^{12}$ yr is
    ruled out by the data.  This constraints the Population~III
    contribution to reionization.  Note that the low-mass IMF models
    are harder to rule out, as discussed in the text.}
  \label{fig:dla_lag}
\end{figure*}

By varying the assumptions about the Population III IMF and the metal
mixing timescale, we have seen that the the self-enrichment timescale
can take on values between $3\times 10^6$--$3\times 10^8$ years,
dramatically impacting the chemical feedback which eventually
terminates Population~III star formation.  These degrees of freedom
result in concomitant uncertainties on the contribution of Population
III star-formaiton to the ionizing photon budget of 1--100\%, since
all of the models were able to match the star-formation and and
reionization observables that we considered.  However, we can
discriminate between these possibilities using accurate chemical
enrichment observations in damped Ly$\alpha$ absorbers (DLAs) at
post-reionization redshifts.  Observations of DLAs can be used to
measure gas-phase metallicities at large cosmological lookback times
with high precision. Furthermore, relative abundances can still be
measured accurately deep into the reionization epoch ($z > 6$) using
metal-line transitions redward of Ly$\alpha$, even though
Gunn-Peterson absorption precludes measurement of neutral hydrogen. In
\citet{2013ApJ...772...93K} we modeled the chemical evolution of DLAs,
and showed how their abundance patterns can be used to constrain
Population~III s cenarios. Here we argue that they can also constrain
the contribution of Population~III stars to reionization.

In our model, we assigned a mass-dependent H~\textsc{i} absorption
cross-section, denoted by $\Sigma$ to each halo in order to predict
the expected distribution of DLA abundance ratios (see
\citealt{2013ApJ...772...93K} for details).  This assignment is
motivated by hydrodynamical simulations \citep{1997ApJ...484...31G,
  2001ApJ...559..131G, 2004MNRAS.348..421N, 2004MNRAS.348..435N,
  2007ApJ...660..945N, 2008MNRAS.390.1349P} and reproduces the
observed DLA metallicity evolution \citep{2012ApJ...755...89R},
incidence rate \citep{2005ApJ...635..123P, 2012A&A...547L...1N}, and
clustering bias \citep{2012arXiv1209.4596F} at low redshifts ($z\sim
3$) very well, and takes the form
\begin{equation}
\Sigma(M)=\Sigma_0\left(\frac{M}{M_0}\right)^2\left(1+\frac{M}{M_0}\right)^{\alpha-2},  
\label{eqn:dlafit}
\end{equation}
where the constants take the values of $\alpha=0.2$, $M_0=10^{9.5}$
M$_\odot$, and $\Sigma_0=40$ kpc$^2$ at $z=3$
\citep{2008MNRAS.390.1349P, 2012arXiv1209.4596F}.  Values at other
redshifts are calculated by mapping haloes at these redshifts to
haloes $z=3$ according to circular velocity
\citep{2012arXiv1209.4596F}.  With this assignment, for any measurable
property $p$ (e.g., abundance ratio [M$_1$/M$_2$]) of DLAs, we can
calculate the number of systems with different values of $p$ in a
sample of DLAs.  This is called the line density distribution, and
with Eqn.~(\ref{eqn:dlafit}) in hand, it can be written as
\citep[e.g.,][]{2005ARA&A..43..861W}
\begin{equation}
\frac{d^2N}{dXdp}=N(M)\cdot\Sigma(M)\cdot\frac{dl}{dX}\frac{dM}{dp}\cdot (1+z)^3.
\label{eqn:d2n}
\end{equation}
Here, $X$ is an absorption length element given by
\begin{equation}
\frac{dl}{dX}=\frac{c}{H_0(1+z)^3},
\label{eqn:dx}
\end{equation}
$dl=cdt$ is a length element, and $p$ is the property in
consideration.  The halo mass is denoted by $M$, $N(M)$ is the
comoving number density of halos (i.e., the halo mass function), and
$\Sigma(M)$ is the halo cross section given by
Eqn.~(\ref{eqn:dlafit}).  The quantity $dM/dp$ in Eqn.~(\ref{eqn:d2n})
can be easily calculated in our model, as properties like metallicity
and relative abundances are known for all halo masses.  The integral
of Eqn.~(\ref{eqn:d2n}) over all values of $p$ is just the total
line density of DLAs, $dN/dX$.  The average value of $p$ in an
observed sample of DLAs is given by
\begin{equation}
  \langle p\rangle=\int\frac{d^2N}{dXdp}\cdot p\cdot dp\cdot dX.
  \label{eqn:abr_avg}
\end{equation}

Fig.~\ref{fig:dla_lag} shows the result of evaluating
Eqn.~(\ref{eqn:abr_avg}) for $p=$ [C/Fe] and [O/Si] in our fiducial
model and its variants.  It also shows the observed evolution of
[C/Fe] and [O/Si] relative abundances in DLAs.  The $z\sim 2$--$4$
measurements are from \citet{2003MNRAS.345..447D, 2007MNRAS.382..177P,
  2011MNRAS.417.1534C} and $z>4$ measurements are from
\citet{2012ApJ...744...91B}, as compiled by
\citet{2012ApJ...744...91B}.  Over a time period of about 6 Gyr (from
$z=2$ to $z=6$), these abundance ratios are relatively constant.
Furthermore, they show little scatter around the mean.  Solid lines in
Fig.~\ref{fig:dla_lag} show the evolution of these relative abundances
in our low-mass IMF models, while the dashed lines show the evolution
in the high-mass IMF case.  Curves with different colors indicate
different delay times.

Our fiducial model agrees with data for both [C/Fe] as well as [O/Si].
In this redshift range, the mean values of these abundance ratios in
this model are governed by the Population~II IMF as the contribution
of Population~III stars is erased in all but the smallest haloes.  The
asymptotic value of these abundance ratios towards low redshift thus
simply reflects the relative yields of these elements per star
integrated over the Population~II IMF.

However, the variants of the fiducial model, in which Population~III
contribution to the cosmic SFR and photoionization rate is higher,
disagree with current relative abunance measurements.  We first
discuss the high-mass IMF models.  As described in the previous
section, we considered two variants where metals are gradually mixed
in the ISM over a period of $t_\mathrm{delay}=10^{11}$ yr and
$10^{12}$ yr respectively, resulting in corresponding self-enrichment
time scales of $10^8$ and $5\times 10^8$ yr.  As shown in
Figs.~\ref{fig:t8} and \ref{fig:t9}, the first of these models has
more than 20\% Population~III contribution to the photoionization rate
down to $z\sim 20$, while in the second model the contribution is more
than 20\% all way down to $z\sim 10$.  We now see in
Fig.~\ref{fig:dla_lag} that the model with $t_\mathrm{delay}=10^{11}$
yr is marginally ruled out by the existing [O/Si] data (red dashed
curve in bottom panel of Fig.~\ref{fig:dla_lag}), while the model with
$t_\mathrm{delay}=10^{12}$ yr is completely ruled out.  In the
previous section we showed that the contribution of Population~III
stars to reionziation could only be enhanced by increasing the
metal-mixing time scale, however this results in a corresponding
increase in the chemical vestiges of these Population~III stars.
Fig.~\ref{fig:dla_lag} shows that the DLA relative abundance
measurements actually restrict the high-mass Population~III
contribution to reionization to be less than 10\% for $z\lesssim 15$.
The model with $t_\mathrm{delay}=10^{11}$ yr acts as a kind of upper
bound on the self-enrichment time scale of haloes and therefore on the
role that high-mass Population~III stars play in hydrogen
reionization.  The [O/Si] ratio is more constraining than [C/Fe]
because of significant variation in the Si yield with IMF: the
high-mass IMF produces two orders of magnitude higher Si yield as Si
is efficiently produced in massive stars due to O-burning.

The contribution of low-mass Population~III stars to reionization is
harder to constrain using DLA chemical data.  This is because chemical
yields of low-mass Population~III stars are not very different from
those of Population~II stars considered in our models, as the two IMFs
have similar shapes and mass ranges.  (There is some difference in the
yields due to difference in their metallicities.)  Thus we see in
Fig.~\ref{fig:dla_lag} that the [O/Si] values in all our low-mass IMF
models are consistent with the data, regardless of mixing delay and
self-enrichment timescale.  This is also true for the [C/Fe] abundance
ratio, although the values are slightly different from the fiducial
case, because the yields are different and the large mixing time scale
slows dilution by Population~II yields.  However, even if we cannot
rule out the low-mass IMF case, it is worth noting that the ionizing
emissvity of low mass Population~III stars is only a factor of two
higher than that of Population~II stars.  Therefore reionization by
these stars is qualitatively similar to reionization by Population~II
stars alone.

Finally, we note that the exact behaviour of the curves in
Fig.~\ref{fig:dla_lag} depends on the form of the mixing function
used.  A mixing function different from that in Eqn.~\ref{eqn:mixfn}
will in general result in a different evolution of the mean relative
abundances.  However, the primary result of Fig.~\ref{fig:dla_lag} is
more general: any Population~III star formation activity that produces
hydrogen-ionizing photons at these redshifts will also necessarily
produce Population~III chemical signatures, which can be constrained
using measurements of abundance patterns in DLAs.  Also DLA relative
abundance measurements at $z\sim 8$, using background QSOs or GRBs,
could start to constrain even the low-mass Population~III IMF by
witnessing the build-up of the metallicity in halos and change in
relative abundances with time because of finite stellar lifetimes.

\section{Conclusion}
\label{sec:discuss}

In this paper, we have demonstrated that Population~III stars
contribute very little to the cosmic SFR density and to the
reionization history.  This is because the halos dominating the cosmic
star-formation rate at high-redshift rapidly (time scales of only
$\sim 10^7$ yr) self-enrich to the critical metallicity, terminating
Population~III star formation.  We quantify this rapid self-enrichment
by defining a halo self-enrichment time scale, which is $\sim 10^6$ yr
in our fiducial model.  This time scale is set by stellar lifetimes,
and is almost independent of halo mass.  Although this rapid
self-enrichment occurs at different redshift in halos of different
masses, the net effect is to reduce the Population~III star formation
and contribution to the hydrogen photoionization rate to less than 1\%
of the total at $z=10$ in our fiducial model.

Previous studies did not uncover this rapid chemical feedback of
Population~III star-formation, because they did not implement a
self-consistent chemical enrichment model, as we have done here.
Since our fiducial model assumes instantaneous mixing of metals in the
ISM, we studied how relaxing this assumption impacts our results.  By
slowing the rate at which metals mix into the ISM, we found that the
termination of Population~III star-formation can be delayed, thus
increasing their contribution to reionization.  However, mixing delay
times and the resulting self-enrichment timescales cannot be
arbitrarily long.  This is because relative metal abundance patterns
in DLAs retain the chemical signatures of Population~III SNe, thus
providing a chemical record of the Population III star-formation
history.  Indeed, we find that halo self-enrichment timescales
significantly longer than $10^8$ yr produce abundance patterns that
are significantly different from those observed in DLAs at $z\lesssim
6$, and are thus ruled out.  As a result, the maximum allowed delay
time implied by existing observations restricts the fractional
contribution of high-mass Population~III stars to the ionization rate
to be $\lesssim 10\%$ at $z=10$.  Constraints on low-mass
Population~III are weaker, because there elemental yields are very
similar to Population~II stars. However, the ionizing emissivity of
low-mass Population~III stars does not significantly differ from
Population~II stars, and so they do little to ease the tension between
reionization constraints and observations of star-forming galaxies at
high redshift.

One possible way in which our chemical constraints can be evaded is by
having an UV photon escape fraction close to 100\% for Population~III
stars, thus dramatically enhancing the impact of Population~III on
reionization for a given amount of Population~III star-formation (and
corresponding heavy element production).  However, in our fiducial
model, the contribution of Population~III stars to the hydrogen
photoionization rate is low predominantly because their contribution
to the cosmic SFR density is extremely low throughout the epoch of
reionization.  As a result, even an escape fraction of 100\% does not
increase their contribution to reionization beyond that of the
Population~II stars.  Additionally, it is not clear what physical
process could lead to such a dramatic increase of the escape fraction
at higher redshifts \citep{2013MNRAS.431.2826F}.

Our work suggests that Population~III stars probably do not resolve
the tension between reionization constraints and the paucity of
ionizing photons implied by the observed population of star-forming
galaxies at high redshift. Looking forward, our model also predicts
that relative abundance measurements in the highest redshift ($z\sim
7$--$8$) QSOs, and possibly also GRBs, should begin to probe the era
when the vestiges of Population~III star-formation had a significant
impact on DLA relative abundance patterns.

\section*{Acknowledgements}

We acknowledge useful discussions with Nishita Desai, Kristian
Finlator, Katherine Inskip, Khee-Gan Lee, J.~Xavier Prochaska, Alberto
Rorai, and Raffaella Schneider. JFH acknowledges generous support from
the Alexander von Humboldt foundation in the context of the Sofja
Kovalevskaja Award. The Humboldt foundation is funded by the German
Federal Ministry for Education and Research.

\end{document}